# Controlled spatiotemporal excitation of metal nanoparticles with femtosecond pulses

Tae-Woo Lee and Stephen K. Gray

Chemistry Division and Center for Nanoscale Materials

Argonne National Laboratory

Argonne, Illinois 60439

## Abstract

We discuss finite-difference time-domain simulations of femtosecond pulses interacting with silver nanowires and nanoparticles.  We show how localized hot spots near the metal surfaces can be generated and controlled in a spatiotemporal manner.   The control is made possible by chirping the pulses such that the effective frequency passes through surface plasmon resonances associated with different spatial regions of the nanostructure over the course of time.   The response of such nanostructures to chirped pulses could provide a novel means of encoding or decoding  optical signals.

## I. INTRODUCTION

The area of coherent control is about using coherent sources of excitation, e.g., lasers, in a variety of ingenious ways to achieve desired outcomes. Particular success in coherently controlling the spectroscopy and dynamics of atoms and molecules should be noted [1-5]. There is also great interest in the coherent excitation and control of a range of other systems, including for example semiconductors [6], quantum dots [7], biological systems [8] and nanostructures [9-17].

There is much current interest in the response of metal nanoparticles [13-15], as well as near-field metallic probes or tips [16,17], to femtosecond or ultrafast pulses. An interesting theoretical demonstration that one of the simpler tools of coherent control, a chirped femtosecond laser pulse, can be used to localize energy in nanostructures was given by Stockman, Faleev, and Bergman [11]. They showed, for example, how such pulses could concentrate energy at the apex of a hypothetical thin V-shaped metallic nanostructure. Motivated in part by such results, we consider here the electrodynamics of larger metallic nanowires and cone-shaped metal nanoparticles interacting with chirped laser pulses.

Our calculations employ the finite-difference time-domain (FDTD) method [18], which is well suited to such time-domain problems. The FDTD method has recently been applied to several related problems [19-22]. We show how the properties of the chirped pulse can be used to control the spatiotemporal behavior of localized excitations or hot spots near the metal surfaces. We discuss how this is due to the creation of localized surface plasmon (SP) excitations [23] near the metal surface.    Counterintuitive, apparent motion of hot spots in directions opposing the propagation direction of the incident light pulses can also be created with appropriate pulses. Such spatiotemporal control, while interesting from a fundamental point of view, could also have practical uses in the areas chemical sensors, optoelectronics and communications. We show, for example,  that the spatiotemporal behavior of surface plasmons in metal nanoparticles can be used to encode or decode digital information in novel ways.

Section II below discusses the metallic nanostructures  of interest to us (IIA),  the laser pulse properties (IIB), and some details of the FDTD calculations (IIC).  Sec. III discusses our results on silver nanowires (IIIA),  cone-like silver nanoparticles (IIIB) and digital information encoding (IIIC).  Sec. IV concludes.

## II. COMPUTATIONAL DETAILS

### A.  Metal nanowire and nanoparticle systems

We consider certain metal nanowire and nanoparticle systems, each of which can be illustrated with the x-z plane diagram of Figure 1.  In the case of the nanowire, we imagine an infinitely long nanowire, surrounded by air,  coming out of the plane of Fig. 1, i.e. invariance of the dielectric properties with respect to y. In this case, Fig. 1 displays the cross section of the nanowire in the x-z plane for any y value. The vertical or z dimension of the nanowire cross section is 250 nm and its horizontal or x dimension varies linearly from 20 nm at its bottom to 140 nm at its top. The second system of interest is a cone-like, cylindrically symmetric metal nanoparticle.   In this case Fig. 1 depicts the x-z plane cross section for y = 0 and the dielectric properties are taken to be invariant with respect to azimuthal rotation about the z axis.  This produces a cone-like particle with a flat bottom or nose.  The nanoparticle is 250 nm high and the diameter of its circular cross section in the x-y plane  varies from 20 nm at its bottom to 140 nm at its top.   This type of structure is also similar to a near-field scanning optical microscope (NSOM) tip [16, 17].   In both cases we consider x-polarized light pulses moving vertically upward along z,  as indicated.

When light interacts with metal nanoparticles it is possible for it to excite collective electronic resonances called surface plasmons (SP's) [23].  SP's can arise if the real part of the metallic dielectric constant, for the applied wavelengths,  takes on certain negative values.  For silver nanosystems these wavelengths are typically in the 300 - 500 nm range.  SP excitations are localized near metal surfaces with electric field intensities much higher than incident intensities.   SP's are also relatively short-lived with characteristic decay times less than 10 fs and are thus broad resonances spanning a range of incident wavelengths.  Note that the SP excitations  we will manipulate here can be best viewed as localized electromagnetic surface waves  and not traveling waves or surface plasmon polaritons [24].  Nonetheless through careful choice of the incident pulse we will be able to create localized excitations that *appear* to move along the surfaces.

The character or field intensity pattern of an SP can vary with wavelength.  It is well known that the peak SP resonance wavelength for metal spheres or cylinders tends to increase with nanoparticle size.  We  might therefore expect that the SP's for a given, less symmetrical structure, to vary in character with wavelength. For example, SP's localized mostly near the narrow, bottom portion of the nanostructure in Fig. 1 might have smaller wavelengths than those mostly localized near the wider, top portion. It is this simple expectation, which turns out to be true, that leads to our ability to obtain spatiotemporal control of intensity in such structures. In particular certain chirped pulses, designed to exhibit different effective frequencies (or wavelengths) at different times, can be expected to excite different spatial regions of the nanostructures at different times.

### B. Femtosecond pulses

We consider incident pulses based on the form [25]

$$f(t) = A(t)\ sin[\omega_0\ t\ + \varphi(t)] \tag{1}$$

where $A(t)$ denotes a pulse envelope function, $\omega_0$ is the carrier frequency, and a linear chirp is introduced by modulating the phase of the sinusoidal carrier function as follows:

$$\varphi(t) = \omega_0\ \alpha\ t^2/\ \tau \qquad . \tag{2}$$

Inserting (2) into (1) and equating the argument of the sine function to $[\omega_{eff}(t)\ t]$ yields the effective frequency

$$\omega_{eff}(t) = \omega_0 \ ( \ 1 + \alpha \ t/\tau) \qquad . \qquad (3)$$

The effective frequency changes linearly from $\omega_0 \ ( \ 1 - \alpha \ )$ to $\omega_0 \ ( \ 1 + \alpha)$ over the duration of the pulse which is taken to be from $t = -\tau \ to + \tau$. A positive chirp, i.e. increasing effective frequency over the duration of the pulse corresponds to $\alpha > 0$ and a negative chirp, i.e. a decreasing effective frequency over the duration of the pulse corresponds to $\alpha < 0$. We use a standard Blackman-Harris window function [26] to define an envelope function $A(t)$ that smoothly changes from 0 to 1 to 0 as $t$ varies from $= -\tau \ to \ 0 \ to \ + \tau$.

Fig. 2 illustrates the pulses produced by Eqs. (1) and (2). For this example we use $\omega_0 = 3.52$ eV ($\lambda_0 = 353$ nm), $\tau = 15$ fs and $|\alpha| = 0.5$. Sub-panels (a) and (b) represent the cases of negative ($\alpha < 0$) and positive ($\alpha > 0$) chirps, respectively. The rather small duration parameter, $\tau$, and large magnitude chirp $|\alpha|$ were chosen to make the chirps easily evident in Fig. 2; more realistic pulses will be considered in our later calculations. It is interesting to note that the energy carried by each pulse is identical in terms of spectral content. Such opposite chirp pulses will be seen to lead to very different spatiotemporal behavior.

## C. FDTD calculations

The finite-difference time-domain (FDTD) method involves simple spatial and temporal discretizations of the time-dependent form of Maxwell's equations, and leads to relatively straightforward direct, time-stepping algorithms [18]. However, the description of metallic regions can be problematic if, as is the case for SP's, the complex, frequency-dependent nature of the dielectric constant is essential. (Even assuming the metallic dielectric constant is real and frequency independent, the fact that it must be negative makes the simplest FDTD algorithm unstable.) However, the following time-dependent form of Maxwell's equations [19],

$$\frac{\partial}{\partial t}\mathbf{E}(x,y,z,t) = \frac{1}{\varepsilon_{\text{eff}}(x,y,z)}\Big[\nabla \times \mathbf{H}(x,y,z,t) - \mathbf{J}(x,y,z,t)\Big]$$

$$\frac{\partial}{\partial t}\mathbf{H}(x,y,z,t) = -\frac{1}{\mu_o}\nabla \times \mathbf{E}(x,y,z,t) \tag{4}$$

$$\frac{\partial}{\partial t}\mathbf{J}(x,y,z,t) = a(x,y,z)\ \mathbf{J}(x,y,z,t) + b(x,y,z)\ \mathbf{E}(x,y,z,t)$$

can be solved in a stable manner with the FDTD method . In parts of space occupied by an ordinary dielectric material with a positive, real and constant relative dielectric constant $\varepsilon_R$, $\varepsilon_{eff} = \varepsilon_R \varepsilon_o$ and $a = b = 0$. In metallic parts of space, if one takes $\varepsilon_{eff} = \varepsilon_\infty \varepsilon_o$, $a = -\Gamma_p$, $b = \varepsilon_o \omega_p^2$, then the solution of (4) is equivalent to using a Drude form [23] for the complex, frequency-dependent metal's dielectric constant,

$$\varepsilon(\omega) = \varepsilon_\infty - \frac{\omega_p^2}{\omega^2 + i\ \Gamma_p \omega} \qquad . \tag{5}$$

Following Ref. [19] we use Drude parameters fit to empirical dielectric constant data for for silver in a specific range of frequencies (or wavelengths). In particular we use $\varepsilon_\infty =$ 8.926, $\omega_p$ = 11.585 eV, and $\Gamma_D$= 0.203 eV, appropriate for describing wavelengths in the range 300 - 500 nm [19].

The symmetry of the metal nanowire case results in a two-dimensional system similar in spirit to that studied in Ref. [19]. With the axis system introduced in Fig. 1 we need only propagate $E_x$, $E_z$ and $H_y$ field components, which in turn depend only on $x$, $z$, and $t$. The relevant FDTD equations are the same form as those in Ref. [19] except that in that reference the infinitely long nanowire axis was taken to be the z-axis instead of the y-axis used here. The computational domain is terminated by a uniaxial perfectly matched layer (UPML) [18] to effectively absorb outgoing field components. The total field/scattered field approach [18] was used to implement the initial chirped pulses, which are given by $E_z(x,z,t) = 0$,

$$E_x(x, z_i, t) = A(t) \; sin[\, \omega_{eff}(t)\, t - k_{eff}(t)\, z_i]$$

$$H_y(x, z_i, t) = (\varepsilon_0/\mu_0)^{1/2}\, A(t)\; sin[\, \omega_{eff}(t)\, t - k_{eff}(t)\, z_i]\;\;, \tag{6}$$

with $A(t)$ and $\omega_{eff}(t)$ as defined in Sec. IIB, and $k_{eff}(t) = \omega_{eff}(t)/c$. Note these pulses are injected into the system, over the initial time period $-\tau$ to $\tau$ defined by $A(t)$, along a given line in $x$ at some fixed $z$ value, $z_i$, outside the nanowire that also serves to separate the total and scattered fields. In practice for the nanowire case, the computational domain consists of 600 x 1000 grid points in the $x$-$z$ plane, with resolution is $\Delta x = \Delta z = 0.5$ nm. and time step $\Delta t = 0.00107$ fs.

The three-dimensional cone-like nanoparticle we also study is is most efficiently modeled with a cylindrical coordinates $(\rho, z, \varphi)$ FDTD algorithm [18, 27]. In this approach, the six field components ($H_\rho$, $H_\varphi$, $H_z$, $E_\rho$, $E_\varphi$, $E_z$) are expressed as general sine and cosine series in terms of $\varphi$. The x-polarization of the initial condition (Fig. 1) allows one to focus only on the components $E^c_\rho$, $E^c_z$ and $H^c_\varphi$ associated with with $\cos(\varphi)$ and the components $E^s_\varphi$, $H^s_\rho$, and $H^s_z$ associated with $\sin(\varphi)$ in such a series. Thus, while one must propagate six field components, computer time and memory are reduced in comparison with a Cartesian coordinates implementation because these components only depend on $\rho$ and $z$. The initial pulse is equivalent to the nanowire case one, i.e., it emanates out of the line $z = z_i$ for all x over the initial $t = -\tau, \tau$ period, with $E^c_z = H^s_z = 0$, and

$$E^c_\rho = E_x \ , \ E^s_\varphi = -E_x$$

$$H^s_\rho = H_y \ , H^c_\varphi = -H_y \quad , \tag{7}$$

with $E_x$ and $H_y$ given by Eq. (6). The computational domain in the nanoparticle case consists of 500 x 900 evenly spaced grid points in the $\rho$-$z$ plane, which is comparable to 1000 x 1000 x 900 points in a Cartesian coordinate system. As with the nanowire case, we use spatial resolution $\Delta\rho = \Delta z = 0.5$ nm, UPML are used to absorb outgoing field components, and the initial pulse was implemented with the total field/scattered field formulation [18]. Owing to a different stability limit [27], the time step is slightly smaller, $\Delta t = 0.00082$ fs.

The computational grids, time steps and other details employed here were designed to yield time-domain intensities converged to better than 5%. We have verified this with convergence checks involving both smaller and larger grid sizes. Each calculation can require up to several hours computation time on a Pentium IV computer.

## III. RESULTS

### A. Nanowire systems

Motivation for determining the pulse parameters in Eqs. (1)-(3), is provided by carrying out preliminary calculations of the frequency response of the silver nanowire (Fig. 1) at positions A and B. Point A is close to the bottom of the nanowire cross section, and point B is close to the top of the nanowire cross section, as indicated in Fig. 1. The absolute square of the Fourier transforms of the electric field at positions A and B, given exposure to an arbitrary femtosecond pulse, shows SP resonance peaks near $\hbar\omega$ = 3.7 eV and 3.4 eV. The chirped pulses are then designed to exhibit effective frequencies spanning this range of frequencies over their durations.

We first consider the nanowire system exposed to chirped femtosecond pulses with $\tau$ = 78.5 fs, central carrier frequency $\hbar\omega_0$ = 3.52 eV, and $\alpha$ = -0.12, 0 and 0.12. The full-width at half maximum (FWHM) such a pulse is 38.7 fs. Fig. 3 shows the envelope of the electric field intensity ($|\mathbf{E}|^2$) observed as a function of time at two different locations, points A and B. The actual field oscillations are so frequent compared to the pulse durations that, for clarity here and from now on, we plot the intensity envelope which covers all the maxima.

Figures 3a, 3b and 3c represent the cases of negative chirp ($\alpha$ = -0.12), no chirp ($\alpha$ = 0), and positive chirp ($\alpha$ = 0.12), respectively. Since all cases use an identical pulse envelope, the *same* amount of  total energy is carried in each pulse.  Several features of the results are worth mentioning. First, controlling electromagnetic energy in a spatiotemporal manner on the sub-wavelength scale can indeed be achieved with the dynamic response of SP's. This is clearly evident with, for example, the negative chirp result showing region A being excited first, followed by B (Fig. 3a) ,  the no chirp case having a more simultaneous excitation of both regions (Fig. 3b)  and, interestingly, the positive chirp case showing B at the far end of the cross section of the nanowire gaining much more intensity than A at first (Fig. 3c). This is consistent with the fact that the resonance frequency at region A is higher than that for region B,  so that time-varying carrier frequency of the negative chirp can pass through the resonance of the region A earlier than that of region B. In positive chirp case,  Fig. 3c, the temporal SP excitations in regions A and B are reversed:  the positive chirp passes through the resonance of the region A *later* than that of region B.

We find the positive chirp case result of Fig. 3c interesting because the actual pulse emanates with positive group and phase velocity (along *z*) from a region below region A,  and thus overlaps region A first and then region B.   However, the temporal progression of the localized spatial excitation  is effectively from region B to region A, implying a negative (in $z$) group velocity. Detailed inspection of movies of the field intensity confirms a picture of positive phase velocity associated with the progression of individual maxima, but with the envelope of several maxima exhibiting a negative group velocity on average. While seemingly counterintuitive, the result is entirely consistent with how the positive chirp has been defined, i.e. as time progresses the effective carrier frequency passes through SP resonances first associated with region B and then region A. The localized and damped nature of the SP excitations also plays a key role. We are not creating traveling surface plasmon polaritons (SPP's). Rather, we are sequentially exciting spatially localized SP resonances along the nanowire surface from the top to bottom in the case of Fig. 3c. Each localized excitation decays in a 1-10 fs timescale, so that newer localized excitations, at more negative z stand out more and lead, on average, to a driven, negative flow of intensity from top to bottom. If we turn off the pulse at any given time, we do not find any significant continuation of the envelope of the exication in a negative $z$ direction, but rather behavior consistent with decay localized SP's.

The intensities shown in Fig. 3 are on a scale such that the maximum intensity of the incident pulse is unity. Intensity enhancements of about an order of magnitude are evident owing to the SP excitations. These enhancement levels are nonetheless smaller than what might be expected. (Somewhat larger intensity enhancements will occur for observation points closer to the metal surfaces.) The reason for this is the finite duration of the laser pulses and the fact that the frequency is never in resonance for very long. This is in contrast to the results of Ref. [11] which showed coherent excitations with much larger intensity enhancements. However, our nanostructures are different both in terms of shape and size than those of Ref. [11] and we also did not attempt to optimize our pulses in any way for intensity enhancement.

Figs. 4 and 5 are nanowire results we have obtained with longer pulse durations: $\tau = 157.0$ fs and $\tau = 628.3$ fs. Otherwise, the same parameters as in Fig. 3 (with $\tau = 78.5$ fs) were used. As with Fig. 3, for clarity just the envelopes of intensity maxima are displayed. By comparing Figs. 3, 4 and 5, corresponding to ever longer pulse durations, it can be seen that the temporal variations of electric field intensity are similar in shape for three cases. However, the temporal delay of SP excitation between regions A and B becomes larger for longer pulses, while the unchirped case still remains as an almost simultaneous excitation. The time separations between the peaks associated with A and B increase approximately linearly with pulse duration. For the negative chirp, one has 19.3 fs (Fig. 3a), 36.8 fs (Fig. 4a), and 141.1 fs (Fig. 5a), and for the positive chirp the separation between regions B and A is 14.7 fs (Fig. 3c), 28.12 fs (Fig. 4c), and 136.3 fs (Fig. 5c). This indicates the temporal scalability of the characteristics of SP excitation in such nanosystems. In other words, the longer pulse width, the more distinctive temporal discrimination of SP excitations in spatial different regions can be obtained. We also conducted simulations for other chirp parameters than those displayed. The results show that the spatiotemporal discrimination becomes weak as the absolute value of the chirp parameter decreases, and eventually converges to the unchirped cases shown above.

The intensity responses of the system scale approximately with $t' = t/(2\tau)$. The time axes in Figs. 3-5 were all chosen to correspond to the same, scaled range $t' = 0.15$ to 0.9. While only clear if one were to overlay the figures, the intensity variation becomes increasingly similar when viewed as a function of $t'$ for increasing pulse length parameters $\tau$.

### B. Cone-shaped nanoparticles

As with the nanowire case above, two observation points A and B are set to compare the temporal variation of local field intensities in two different locations near the nanoparticle (Fig. 1). If $(x_A, z_A)$ denotes observation point A in the nanowire case above, then we use observation point $(\rho_A, z_A, \varphi_A) = (|x_A|, z_A, 0)$. We also carried out a preliminary frequency response analysis at points A and B, inferring the SP resonance frequencies near $\hbar\omega = 3.55$ eV and 3.4 eV, similar to the nanowire case. We chose to study central pulse frequency and width $\hbar\omega_0 = 3.47$ eV and $\tau = 628.3$ fs, as well as chirps $\alpha = -0.12$, 0 and 0.12.

Fig. 6 shows the intensity envelope variation in time for regions A and B for the three different chirp cases. The results are very similar to those obtained for the nanowire case with a similar pulse, Fig. 5. Thus, the negative chirp (Fig. 6a) leads to clear excitation of region A followed by region B, the no chirp case (Fig. 6b) has almost simultaneous excitation of the regions and the positive chirp shows region B being excited prior to region A. We also conducted the FDTD simulations for different pulse widths. The results follow what we have seen in nanowire case. The spatiotemporal discrimination of the SP excitation is again almost linearly proportional to the width in time of the incident pulse, indicating good temporal scalability. The peak differences in time are 12.8 fs, 30.3 fs, and 131.5 fs for negative chirp and 14.0 fs, 30.0 fs, and 131.0 fs for positive chirp when incident pulse widths are $\tau=$ 78.5, 157.0, and 628.3 fs, respectively. Thus we have shown that there is nothing special about the more explicitly 2-D nanowire configuration and that cone-like 3-D nanoparticles also display very similar spatiotemporal behavior.

### C. Digital information from optical pulses

We have seen in the above two subsections that suitably chosen, chirped optical pulses can generate hot spots on certain metal nanosystems with controllable spatiotemporal behavior. An interesting twist on this is to use such spatiotemporal evolution to generate a new or more desirable signal than the originally applied one. Such manipulations could be of use in optoelectronics and communications [25, 28]. In conventional optical data communication systems, for example, digitally formatted information is transferred along optical waveguides as pulsed optical signals. Since each pulse (or its absence) represents a digital signal of 1 (or 0) in a given clock cycle, it is essential to avoid overlap of adjacent pulses. This condition becomes one of the obstacles in using full capacity of the wide bandwidth of ultra-short pulses or chirped pulses. (Chirped pulses have larger bandwidth than unchirped pulses with same pulse duration.) In the chirped pulse case, the main factor limiting bit-rate is not the bandwidth of the pulses but delay time for the sequence of pulses which must not overlap. This is also true for ultra-short pulses since chirped pulses can be considered as a result of ultra-short pulses spread out by group velocity dispersion. However, as we will demonstrate, nanosystems can decode the original digital information from severely overlapping pulses, suggesting a means of increasing the bit-rate. Alternatively, one might wish to encode the digital information from the start with less obviously decodable overlapping pulses.

As a concrete example, imagine a digital system with a 150 fs clock cycle. A digital sequence (1,0,1) could be represented with an initial pulse spanning an approximate FWHM much less than 150 fs and another, similar pulse separated from the original by 300 fs. On the basis of intensities, over the first 150 fs one would read '1', over the next 150 fs one would read '0' and over the final 150 fs one would read '1'. Suppose however the pulse sequence over the time period in question was given by Fig. 7a, which corresponds to the envelope of the intensity maxima of two overlapping, negatively chirped pulses. Each chirped pulse has parameters $\alpha$ = -0.12, $\omega_0$ =3.47 eV, and $\tau$ =561 fs (Sec. IIB), with the duration parameter $\tau$ chosen such that when displaced by 300 fs in time there is significant overlap. The intensity envelope in Fig. 7a exhibits very high frequency oscillations in the middle time range owing to the superposition of different effective frequencies. Such high frequency oscillations are not easily detected due to the finite response time of photodectors. A local time average of the intensity would be more representative of a measured signal. Fig. 7b displays such a local time-average based on a Gaussian window about each time point with 1/e full-width of 20 fs. While Fig. 7b is now more consistent with the expected (1,0,1) string of digital information, the '0' corresponds to a very shallow minimum that might not be clearly discerned from '1' in practice.

We now assume we are presented with the pulse sequence in Fig. 7 and wish to deduce an unambiguous digital signal. Suppose this pulse sequence is exposed to the 3-D cone-shaped metal nanoparticle in the manner indicated by Fig. 1 and discussed in Sec. IIIB above. (We could equally discuss the nanowire case.) We carried out the corresponding FDTD calculations with the pulse sequence of Fig. 7 as source. During the simulation, we calculated the time-dependent intensities at regions A and B just as before. Fig. 8a displaces the resulting local time averaged intensities. The intensity does not follow the incident pulse shape and the SP excitation in region B is delayed in time relative to region A. Since Fig. 8a shows the local time-averaged field intensities near regions A and B, it is also an approximate representation of an electric signal that could be obtained from the detection of SP's using, e.g., photodiodes designed to register fields near the corresponding regions. Once the optical signals are converted to electric signals (voltages), the signal from B can be *subtracted* from the A signal. Figure 8b shows the resulting difference signal. We define a '1' to be, within the given clock cycle, a predominantly positive-valued difference signal with a pronounced maximum. A '0' is defined to be a predominantly negative-valued signal. With these definitions, Fig. 8b readily yields the correct (1,0,1) digital sequence. Note also that the delay time between maxima of the difference signal is 300 fs, which now agrees well with original pulse delay time. Thus, for a given digital clock cycle, the nanoparticle system successfully decodes severely overlapping optical signals.

If one were one to feed the nanoparticle system a well separated sequence of pulses, a pattern of maxima consistent with the original signal will result upon subtraction of signal B from signal A. One can infer from Fig. 6a, for example, that the difference signal for an isolated pulse would have just one positive-valued maximum over its cycle, i.e. would be interpreted to be a '1' as it should be. Thus the nanoparticle system will still produce the correct result if the pulses are not overlapping.

## IV. CONCLUDING REMARKS

In this paper we studied, with a realistic time-domain computational electrodynamics approach, how hot spots near the surfaces of certain metal nanowires and nanoparticles could be generated with chirped femtosecond pulses. Remarkable spatiotemporal control could be achieved, including counterintuitive motion of hot spots against the light propagation direction. This control was made possible by designing the chirp such that the effective frequency of the pulse resonates with different SP excitations located along the nanostructure over the duration of the pulse. Along with the pioneering theoretical demonstration of Stockman, Faleev and Bergman [11] that localized coherent excitation of small metallic nanostructures is possible, we hope that such results will stimulate new theoretical and experimental explorations of such possibilities.

We also discussed how the response of such nanosystems to chirped pulses could be used to encode or decode digital information. It is also possible that such ideas could be of relevance to chemical sensing. For example, one can imagine applying a chirped pulse such that one knows certain regions of the nanoparticle, at certain times, are excited. The spectral response of molecules near such hot spots should be enhanced (e.g., surface-enhanced Raman spectroscopy [29, 30]). Applying various time-delayed probe pulses could then lead to transient spectral responses indicative of whether or not molecules are near a given place on the nanoparticle.

The analysis presented here was based on simple, linearly chirped pulses. While we did explore some features due to two (overlapping) pulses, it would certainly be interesting to study how spatiotemporal control can further influenced by consideration of two or more pulses, as well as more general pulse shapes.

**ACKNOWLEDGEMENTS**

We thank Gary Wiederrecht for many helpful discussions. This work was supported by the U. S. Department of Energy, Office of Basic Energy Sciences, Division of Chemical Sciences, Geosciences, and Biosciences under DOE contract W-31-109-ENG-38.

**REFERENCES**

[1]     D. J. Tannor and S. A. Rice, Adv. Chem. Phys. **70**, 441 (1988)

[2]     P. Brumer and M. Shapiro, Annu. Rev. Phys. Chem. **43**, 257 (1988).

[3]     W. S. Warren, H. Rabitz, and M. Dahleh, Science **259**, 1581 (1993).

[4]     H. Kawashima, M. M. Wefers, and K. A. Nelson, Annu. Rev. Phys. Chem. **46**, 627 (1995).

[5]     T. Brixner, N. H. Damrauer, and G. Gerber, in *Advances in Atomic, Molecular, and Optical Physics*, Volume 46, pp. 1-54, edited by B. Bederson and H. Walther, (Academic Press, San Diego, 2001).

[6]     J. Erland, V. G. Lyssenko, and J. M. Hvam, Phys. Rev. B **63**, 155317 (2001).

[7]     X. Q. Li, Y. W. Wu, D. Gammon, T. H. Stievater, D. S. Katzer, D. Park, C. Piermarocchi, and L. J. Sham, Science **301**, 809 (2003).

[8]     J. L. Herek, W. Wohlleben, R. J. Cogdell, D. Zeidler, and M. Motzkus, Nature **417**, 533 (2002).

[9]     J. H. Hodak, A. Henglein, and G. V. Hartland, J. Phys. Chem. B 104, 9954 (2000).

[10]    M. I. Stockman, Phys. Rev. Lett. **84**, 1011 (2000).

[11]    M. I. Stockman, S. V. Faleev, and D. J. Bergman, Phys. Rev. Lett. **88**, 067402 (2002).

[12]    K. Komori, G. Hayes, T. Okada, B. Deveaud, X. L. Wang, M. Ogura, M. Watanabe, Jpn. J. Appl. Phys. **41**, 2660(2002).

[13]    S. Link and M. El-Sayed, Annu. Rev. Phys. Chem. **54**, 331 (2003).

[14]    J. Lehmann, M. Merschdorf, W. Pfeiffer, A. Thon, S. Voll, and G. Gerber, Phys. Rev. Lett. **85**, 2921 (2000).

[15]    J. H. Hodak, A. Henglein, and G. V. Hartland, J. Phys. Chem. B **104**, 9954 (2000).

[16]    E. J. Sanchez, L. Novotny, and X. S. Xie, Phys. Rev. Lett. **82**, 4014 (1999).

[17]    A. Pack, M. Hietschold, and R. Wannemacher, Ultramicroscopy **92**, 251 (2002).

[18]    A. Taflove and S. C. Hagness, *Computational Electrodynamics: The Finite-Difference Time-Domain Method, 2nd ed.* (Artech House, Boston, 2000).

[19]    S. K. Gray and T. Kupka, Phys. Rev. B **68**, 045415 (2003).

[20]    J. M. Oliva and S. K. Gray, Chem. Phys. Lett. **379**, 325 (2003).

[21]    G. A. Wurtz, J. S. Im, S. K. Gray, and G. P. Wiederrecht, J. Phys. Chem. B **107**, 14191 (2003).

[22]    L. Yin, V. K. Vlasko-Vlasov, A. Rydh, J. Pearson, U. Welp, S.-H. Chang, S. K. Gray, G. C. Schatz, D. E. Brown, and C. W. Kimball, Appl. Phys. Lett., *submitted* (2004).

[23]    C. F. Bohren and D. R. Huffman, *Absorption and Scattering of Light by Small Particles*, (Wiley, New York, 1983).

[24]    A. V. Zayats and I. I. Smolyaninov, J. Opt. A: Pure Appl. Opt. **5**, S16 (2003).

[25]    A. Yariv, *Optical Electronics in Modern Communications, Fifth Ed.,* (Oxford Univ. Press, New York, 1997).

[26]    A. H. Nuttall, IEEE Trans. on ASSP **29**, 84 (1981).

[27]    D. W. Prather and S. Shi, J. Opt. Soc. Am. A **16**, 1131 (1999). N.b. there are some typographical errors in Eqs. (4)-(10).

[28]    A. Rogers, *Essentials of Optoelectronics with Applications*, (Chapman and Hall,

New York, 1997).

[29]   A. Wokaun, Mol. Phys. **56**, 1 (1985).

[30]   S. Nie and S. R. Emory, Science **275**, 1102 (1997).

**FIGURE CAPTIONS**

**Figure 1.** Diagram of the nanowire and nanoparticle geometries considered. The grey area corresponds to silver. In the nanowire case, the nanowire is imagined to be infinitely extended along the y-axis normal to the x and z axeses. In the nanoparticle case, symmetry with respect to rotation about the z-axis defines a cone-like metal nanoparticle with a blunt lower nose. Points A and B near the bottom and top of the nanostructure are used in our analysis. Specifically, point A is 11.25 nm along $z$ above the bottom of the nanowire, and 2 nm along $x$ (or $\rho$) direction from the nanowire surface. Point B is 238.75 nm along $z$ from the bottom of the nanowire, and also 2 nm along $x$ direction away from the surface.

**Figure 2.** Negative, (a), and positive, (b), frequency-chirped pulses. The pulses shown are somewhat shorter and more strongly chirped than the more experimentally achievable pulses that we employ in our detailed numerical work in order to show clearly the qualitative form of the chirped pulses.

**Figure 3.** Time-evolving intensity envelope at points A and B as the nanowire system in Fig. 1 interacts with the negative chirp, (a), no chirp, (b), and positive chirp cases. The chirp magnitude is $|\alpha| = 0.12$ and duration parameter $\tau = 78.5$ fs.

**Figure 4.** Time-dependent intensity enevelopes at points A and B in the nanowire system. Similar to Fig. 3 except that the pulse duration parameter is $\tau$ = 157.0 fs.

**Figure 5.** Time-dependent intensity envelopes at points A and B in the nanowire system. Similar to Fig. 3 except that the pulse duration parameter is $\tau$ = 628.3 fs.

**Figure 6.** Time-dependent intensity envelopes at points A and B as the cone-shaped nanoparticle system interacts with the negative chirp, (a), no chirp, (b), and positive chirp cases with chirp magnitude $|\alpha|$ = 0.12 and duration parameter $\tau$ = 628.3 fs.

**Figure 7.** (a) The time-dependent intensity envelope of a pulse sequence corresponding to two overlapping, chirped pulses. (b) The averaged intensity of the pulse sequence in (a).

**Figure 8.** (a) Averaged intensities for regions A and B of a cone-shaped silver nanoparticle exposed to the pulse of Fig. 7. (b) The difference of averaged intensities for A and B.

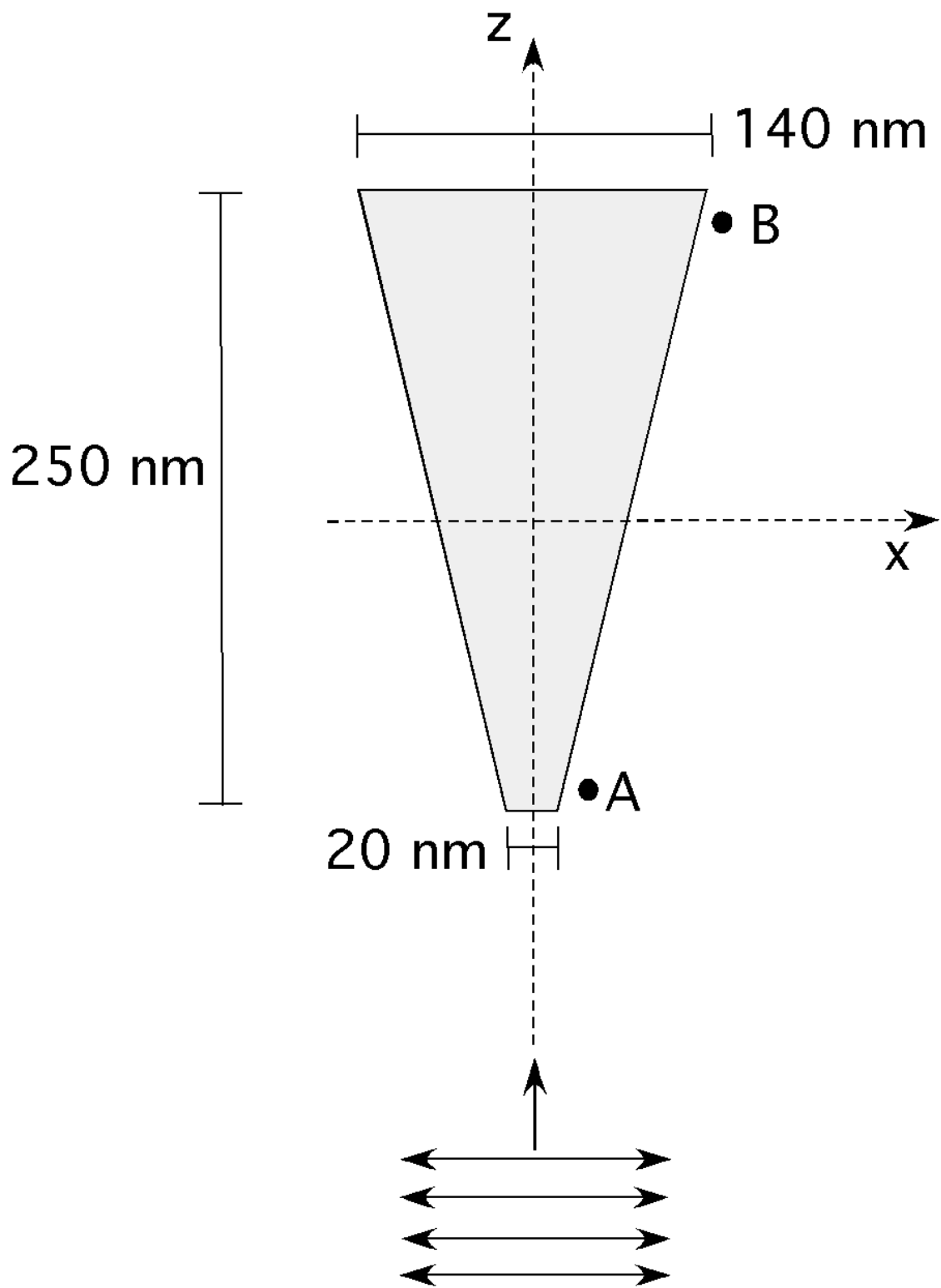

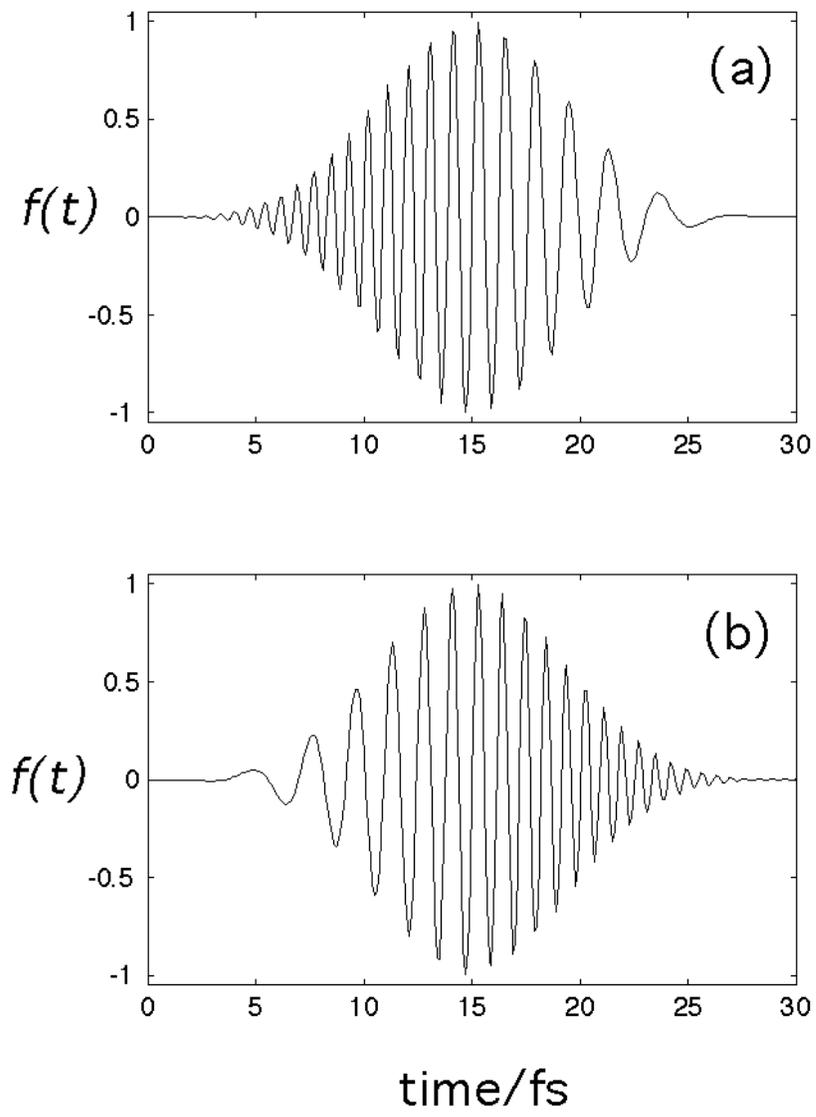

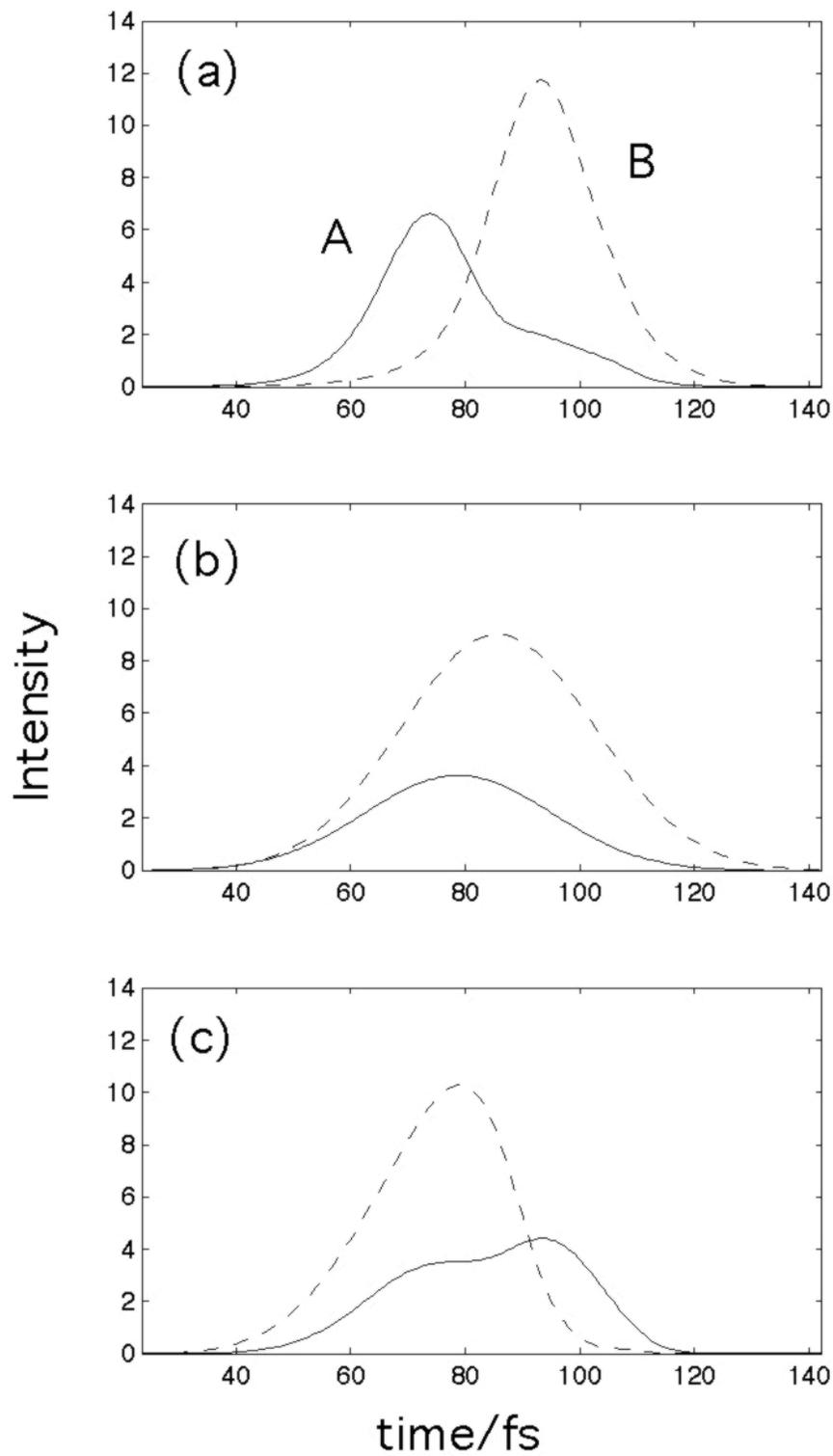

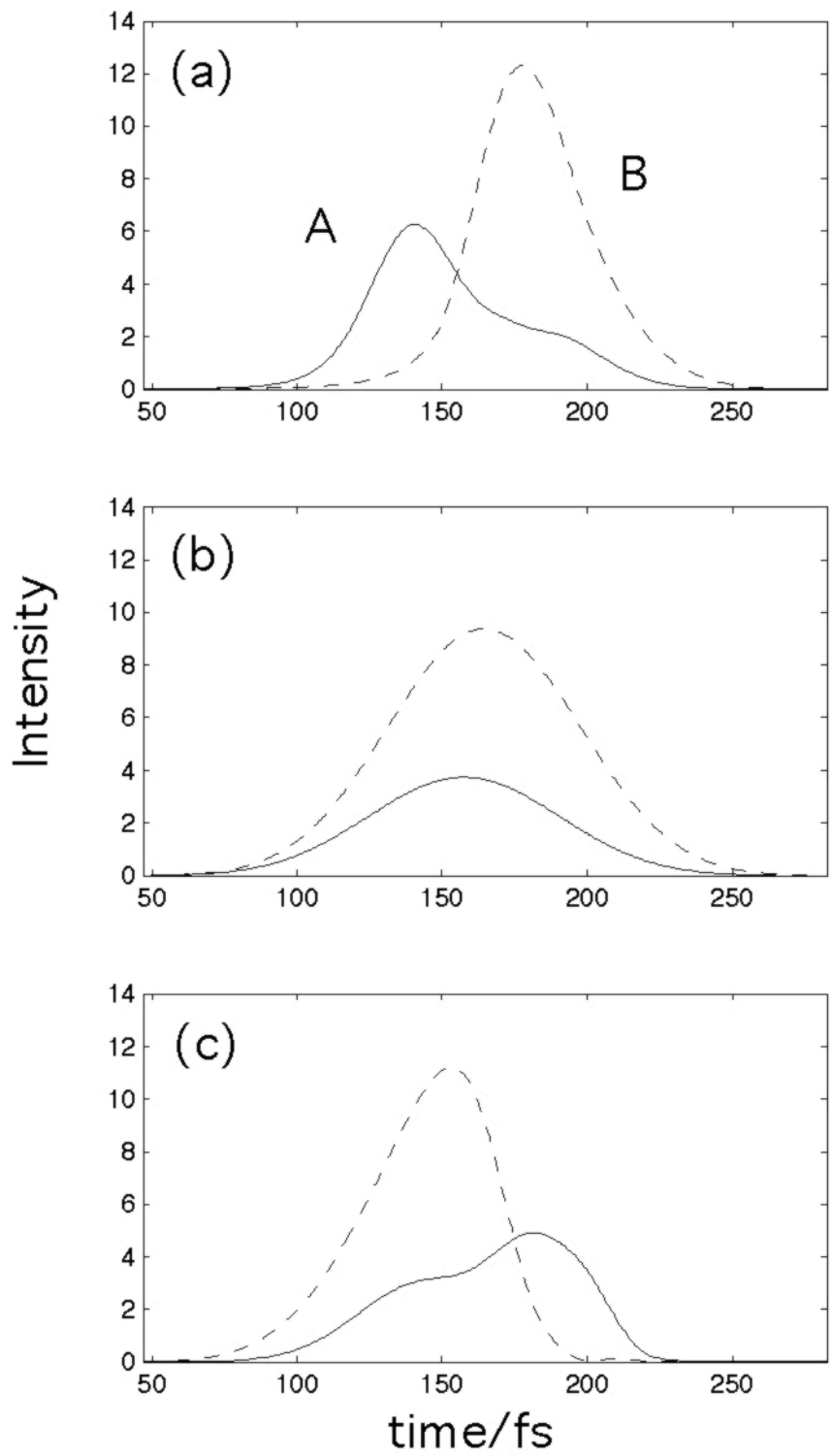

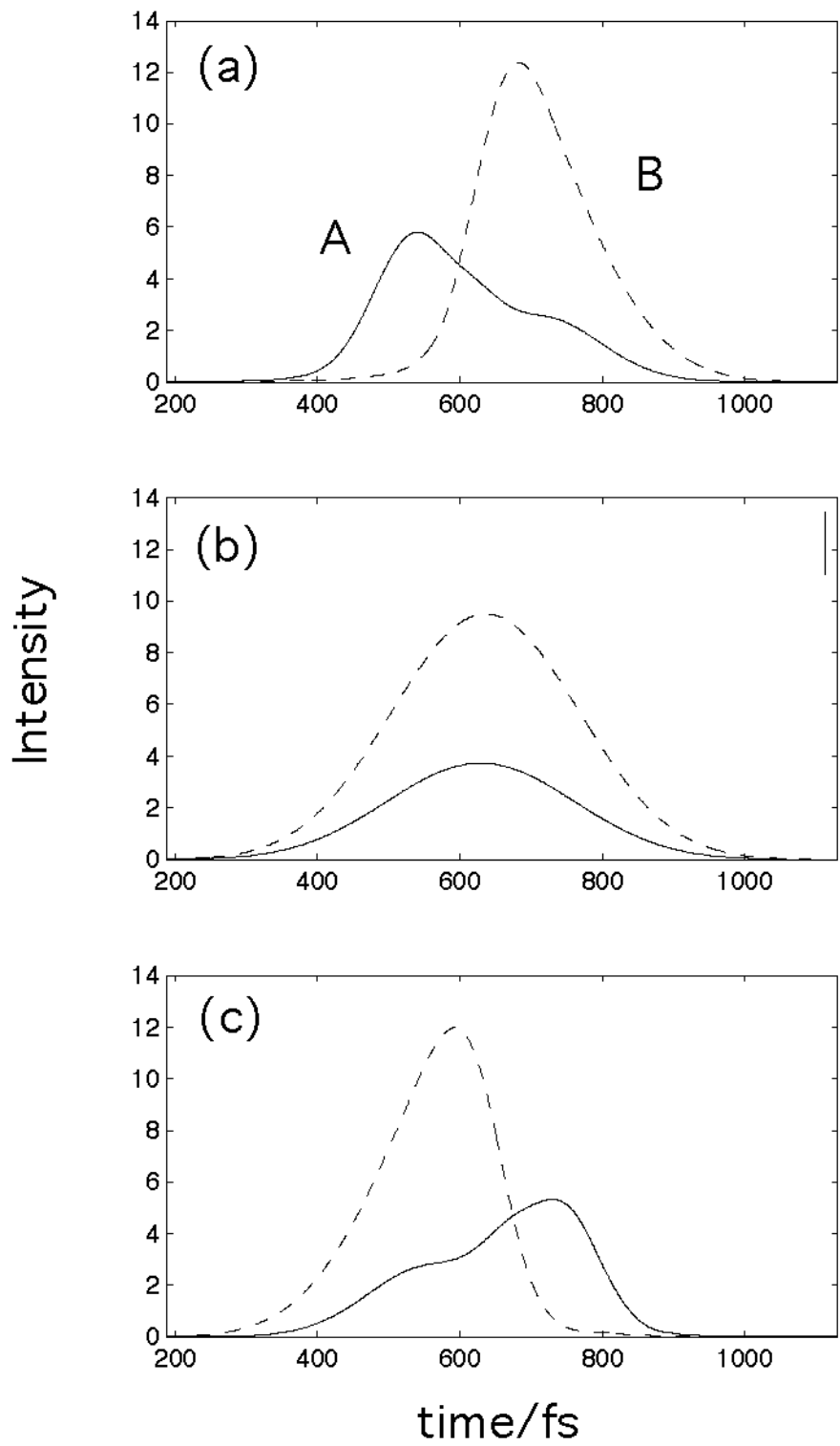

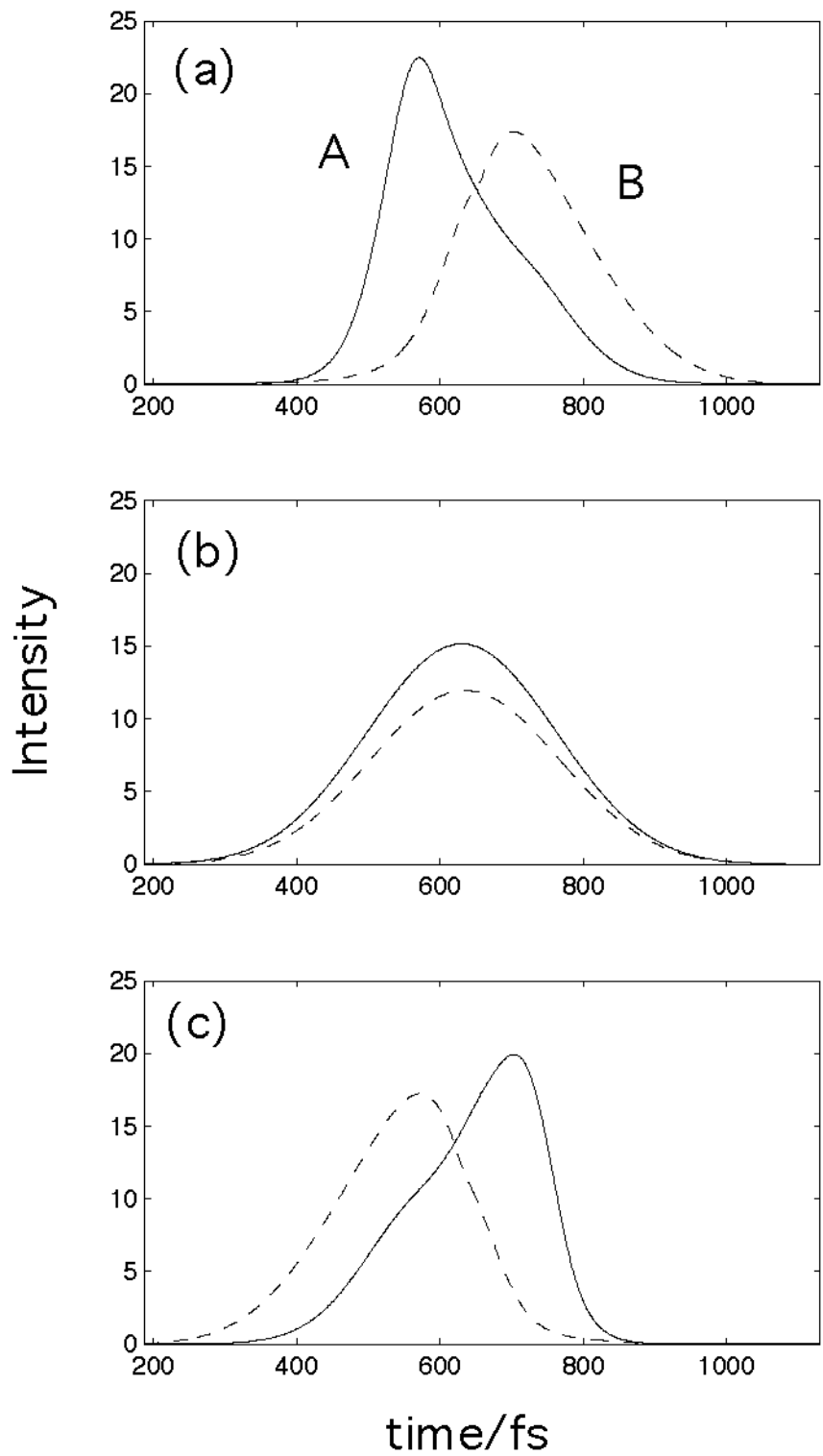

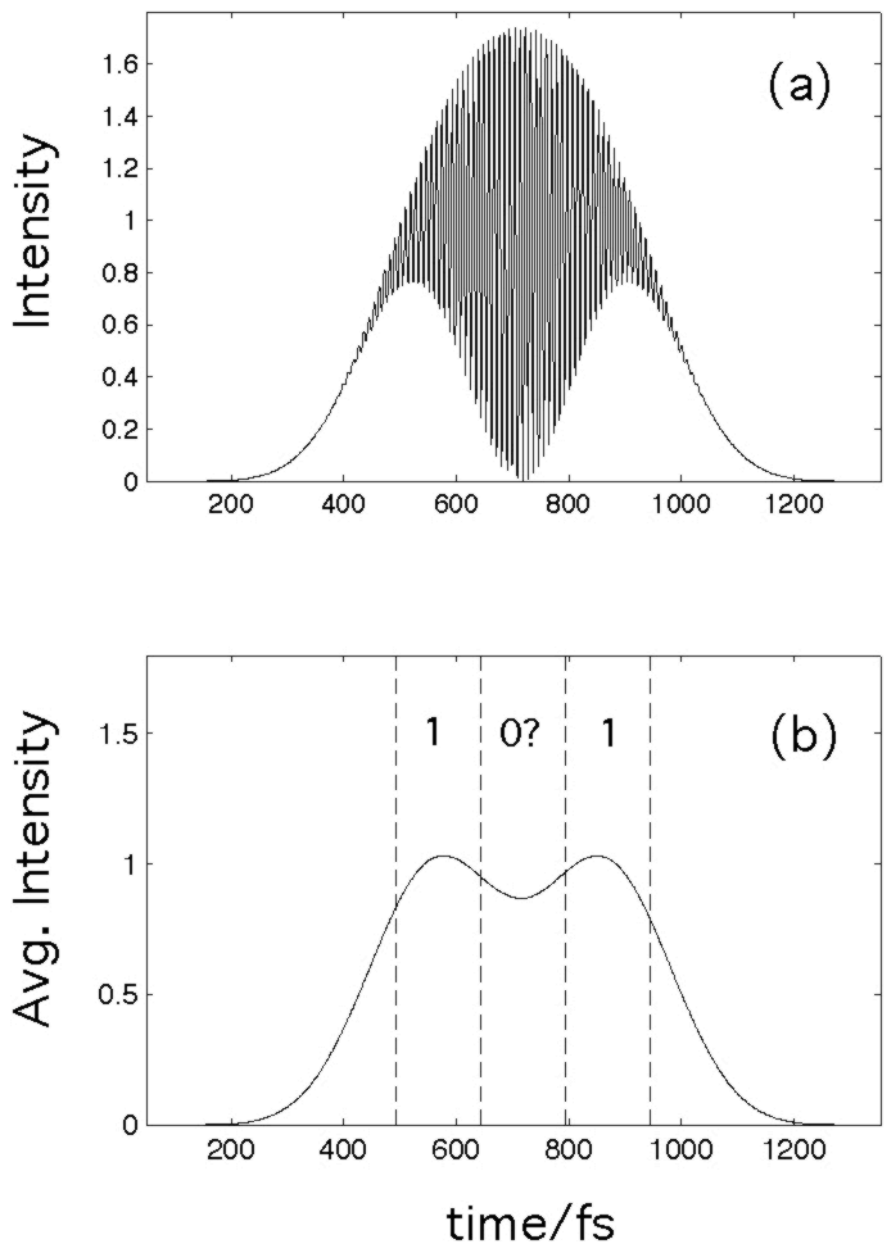

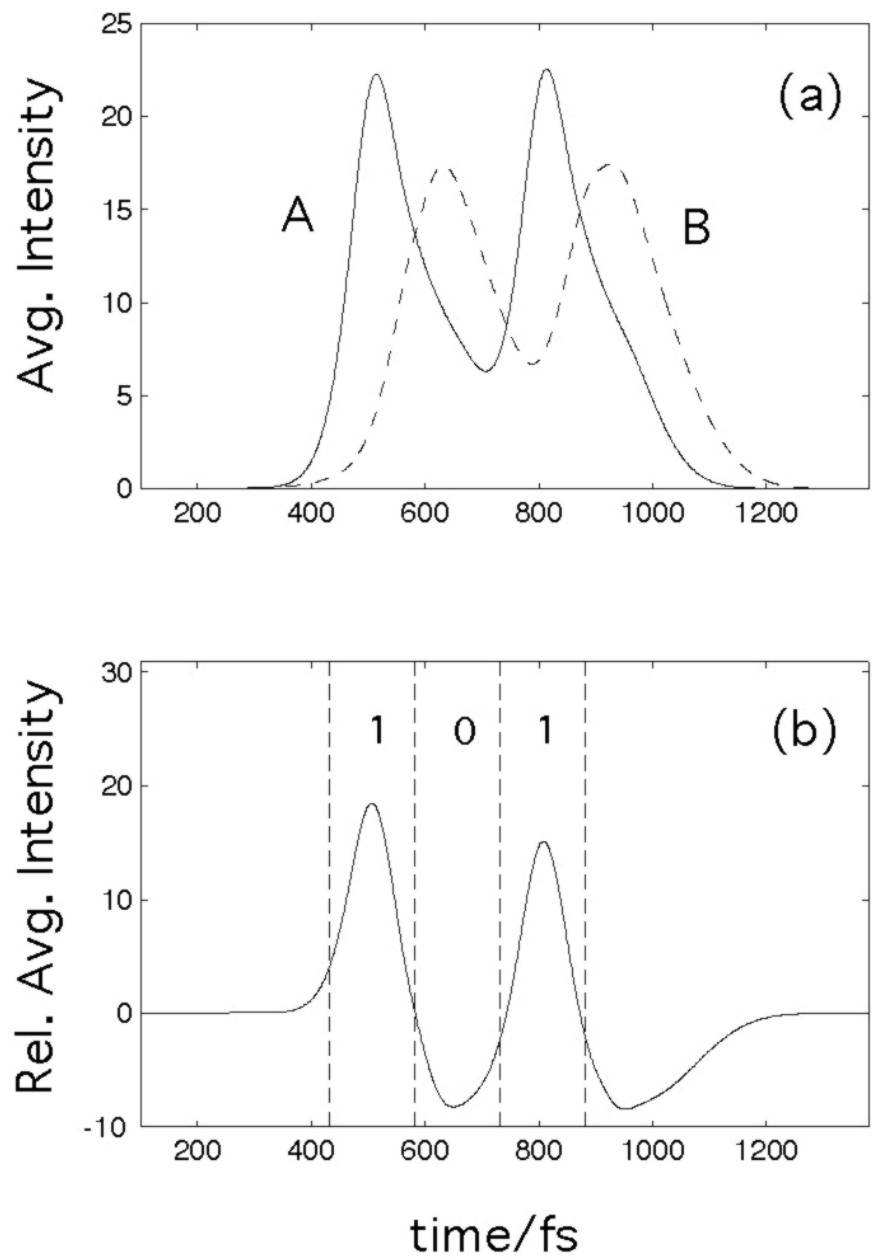